\documentclass[referee]{raa}            

\usepackage{graphicx,times}             
\usepackage{natbib}
\usepackage{amssymb,amsmath}
\bibpunct{(}{)}{;}{a}{}{,}

\usepackage[pagebackref=true]{hyperref}

\begin{document}

  \title{Mock Observations for the CSST Mission: End-to-End Performance Modeling of Optical System}

   \volnopage{Vol.0 (20xx) No.0, 000--000}      
   \setcounter{page}{1}          

   \author{Zhang Ban
      \inst{1}
   \and Xiao-Bo Li
      \inst{1}
   \and Xun Yang
      \inst{1}
   \and Yu-Xi Jiang 
      \inst{1}
   \and Hong-Cai Ma  
      \inst{1}
         \and Wei Wang  
      \inst{1}
        \and Jin-guang Lv  
      \inst{1}
   \and Cheng-Liang Wei 
      \inst{2}
   \and De-Zi Liu 
      \inst{4}
   \and Guo-Liang Li 
      \inst{2}
   \and Chao Liu 
       \inst{3}
   \and Nan Li 
       \inst{3}
    \and Ran Li 
       \inst{3,5}
    \and Peng Wei 
       \inst{3}      
    }

   \institute{Space Optics Department, Changchun Institute of Optics, Fine Mechanics and Physics, Chinese Academy of Sciences, Changchun 130033, China; {\it lixiaobo104@163.com}\\
        \and
             Purple Mountain Observatory, Chinese Academy of Sciences, Nanjing 210023, China.
        \and
             National Astronomical Observatories, Chinese Academy of Sciences, Beijing 100101, China.\\
        \and
             South-Western Institute for Astronomy Research, Yunnan University, Kunming 650500, China.\\
        \and School of Physics and Astronomy, Beijing Normal University, Beijing 100875, China
            \\
\vs\no
   {\small Received 20xx month day; accepted 20xx month day}}

\abstract{This study presents a comprehensive end-to-end simulation analysis of the optical imaging performance of the China Survey Space Telescope (CSST) under in-orbit conditions. An integrated system model incorporating five static and two dynamic error sub-models was established. Wavefront errors were calculated for each sub-model and compared to the integrated system error to quantify the individual contributions to image degradation. At the detector level, wavefront error, point spread function (PSF), and ellipticity were evaluated across the full field of view (FOV). The average radius of 80\% encircled energy (REE80) of the PSF under full-error conditions was determined for 25 field points, yielding a value of 0.114 arcseconds. Furthermore, the calculations indicate a correlation between the wavefront distribution and the ellipticity distribution within the optical system. By optimizing the wavefront distribution, it is possible to adjust the ellipticity distribution of the PSF across the full FOV. The end-to-end simulation approach adopted in this paper provides a theoretical foundation for improving the image quality in large-aperture, off-axis space telescopes.  
}

   \authorrunning{Zhang Ban et al }            
   \titlerunning{Mock Observations for the CSST Mission: End-to-End Performance Modeling of Optical System}  

   \maketitle
%
\section{Introduction}
The CSST is an off-axis three-mirror space telescope designed for a space-based sky survey. Its primary goal is to provide high-resolution astronomical images that will advance the study of galaxy formation and evolution and help constrain cosmological models\citep{2021Zhan}.

Space telescopes are vital for capturing astronomical images, yet assessing their in-orbit performance through direct testing is challenging. Consequently, simulation methods become essential in evaluating imaging performance. Accurate simulations not only help to understand the characteristics of imaging errors \citep{2020O.B.Kauffmann1,2020S.Pires,2021Nan_Li}, but also assist in shortening development time and reducing development costs while facilitating system optimization. Previous research has demonstrated that the integrated opto-mechanical-thermal simulation analysis method can comprehensively assess the imaging performance of space telescopes. Moreover, this method can identify the key factors that contribute to the degradation of image quality  \citep{2012C.Bonoli,2014A.J.Connolly}. The end-to-end system simulation method extends the integrated opto-mechanical-thermal approach. By modeling various error sources through dedicated sub-models based on the physical imaging mechanisms of the telescope, this method computes the system’s imaging performance parameters \citep{2005M.W.Fitzmaurice1a}. The end-to-end simulation analysis supports the entire lifecycle of a space telescope, from design and development to in-orbit operation and post-deployment maintenance.  

This approach has been widely adopted in major space telescope projects. For example, the Hubble Space Telescope (HST) team used it to calculate the PSF of various detectors under in-orbit working conditions and developed the simulation software, Tim Tiny, which can analyze the imaging performance of the telescope\citep{2011J.E.Krista,2007J.D.Rhodes,2020A.Gáspára}. Similarly, the Spitzer Infrared Space Telescope team used it in assessing the optical performance of the telescope under low-temperature operating conditions \citep{2004R.D.Gehrz}. The James Webb Space Telescope (JWST) team adopted it to simulate the dynamic PSF of the coaxial optical system \citep{2023M.W.McElwain}. Furthermore, the European Athena Space Telescope mission team employed it to predict in-orbit imaging performance, providing valuable insights for optimizing and designing subsequent products \citep{2019T.Dauser}.

During the research and development phase of the CSST telescope, end-to-end optical image quality simulations were conducted based on the design parameters, including calculating both the telescope’s PSF and field distortion characteristics. The resulting data were integrated into simulation workflows for all CSST terminal modules to validate the system-level performance. While available space telescopes, such as the HST and JWST, predominantly utilize coaxial optical designs optimized for deep-field observations, our investigation concentrated on off-axis three-mirror anastigmatic (TMA) configurations. Additionally, errors that occur during the operational phase of space telescope optical systems were analyzed comprehensively, including processing errors, environmental deformations, misalignments, and vibrations. In this way, the fidelity of our computational results was enhanced, making them more reflective of real-world conditions.

In summary, the end-to-end simulation modeling process of the CSST optical system was methodically outlined in this paper, and the optical imaging quality characteristics were calculated under this framework. The detailed contents are organized as follows: Section 2.1 presents our static error model; Section 2.2 describes our dynamic error model; Section 2.3 introduces the sampling model at the detector end; Section 3 provides an image quality analysis, and Section 4 concludes the study.

\section{Simulation Modeling}
This section elaborates the modeling framework for the end-to-end simulation of the optical system. During the initial design phase, the imaging capability of the telescope optical system, as described in this paper, approached the diffraction limit. However, as construction proceeded, the imaging performance becomes compromised due to factors such as material processing and assembly adjustments. In the operational phase, the system’s imaging performance further degrades due to the combined effects of environmental variations and self-induced vibrations. These influencing factors do not counterbalance each other; instead, they act simultaneously, resulting in a progressive decline in the imaging performance of the telescope optical system.

First, we analyzed the error factors present during the in-orbit phase of the telescope and establish an error model. The error model is classified into two categories: static and dynamic. Fig. 1 shows a schematic diagram of the end-to-end simulation model for the telescope’s optical system. The central purple dashed box represents the optical system design model, which, in combination with the four error models shown in the left purple dashed box, forms the static error models. The upper-middle purple dashed box denotes two dynamic error models, while the lower-middle purple dashed box represents the active optics model. The pink solid boxes categorize optical system error parameters into mirror surface deformation error parameters and mirror positional error parameters. The right section of the diagram includes a blue dashed box representing image quality results and a brown dashed box representing image quality analysis. Hollow arrows indicate the generation or transfer of intermediate data. Next, we will introduce the modeling process for each of the error models in detail.

\begin{figure}[htbp]
    \centering
    \includegraphics[width=1\linewidth]{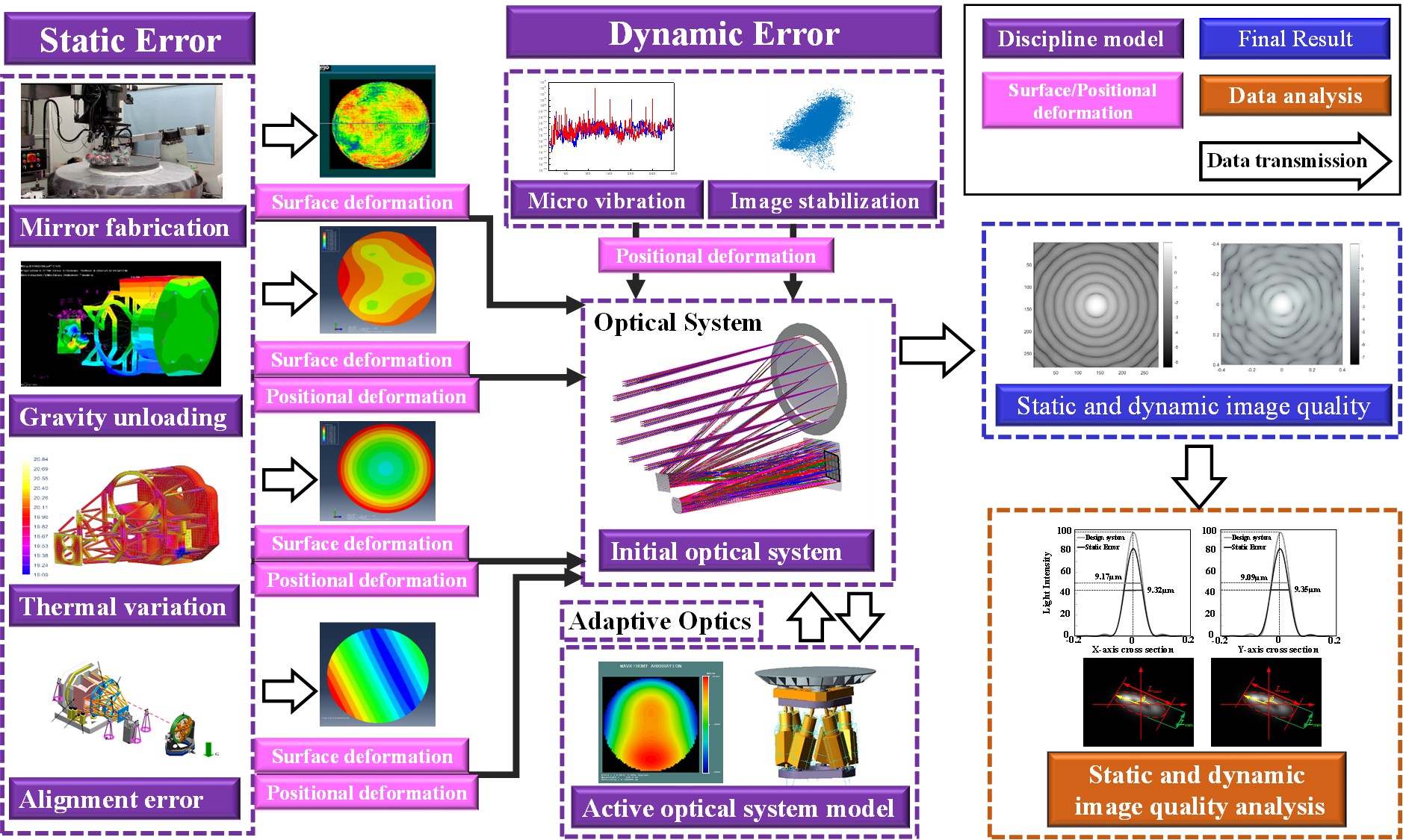}
    \caption{Schematic diagram of the optical system end-to-end simulation modeling.} \label{Fig1}
\end{figure}
\subsection{Static error \textbf{model} }
The static error model consists of five models, namely the optical system design residuals, mirror surface figure error, alignment error, in-orbit gravity release residual, and in-orbit environmental change error, as detailed in following sections.
\subsubsection{\textbf{Optical System Design Residual} }
\begin{figure}[htbp]
    \centering
    \includegraphics[width=0.6\linewidth]{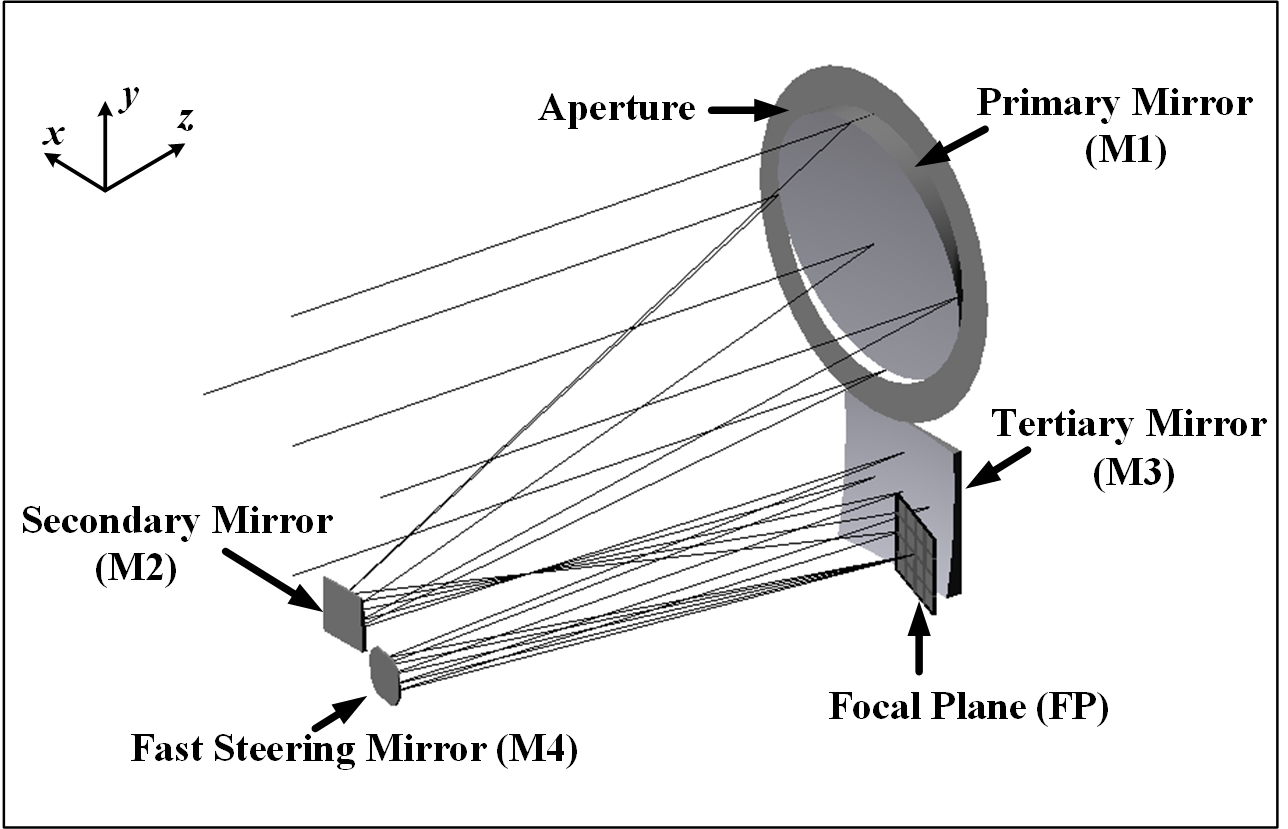}
    \caption{Schematic diagram of an off-axis three-mirror optical system. In the upper-left corner of the diagram, the three arrows denote the coordinate system of the optical system, while the single arrows indicate the labels of individual components within the optical system.}
    \label{Fig2}
\end{figure}

All optical systems contain optical design residuals, which are typically calculated directly using optical design software such as CODE V or Zemax. An off-axis three-mirror reflective system was selected as the optical system in this paper, as illustrated in Fig. \ref{Fig2}. The system comprises four mirrors, labeled M1 to M4, with a F-number of 14, an aperture diameter of 2 meters, and an effective FOV of 1.1°×1°. This paper primarily focuses on the end-to-end simulations based on the optical system design models, and the design residuals were calculated by referring to the optical design-related literature\citep{2014Zeng}. Under the initial design conditions, the wavefront aberration of the optical system is calculated as 0.027$\lambda$ .

The active optics model, serving as a sub-model within the end-to-end system simulation framework, is only activated only when the telescope is in a degraded imaging state during in-orbit operations. However, this process does not fall in the scope of the current study. A detailed description of the operational workflow of the system is given by Ju\citep{2017Ju}.

\subsubsection{\textbf{Mirror Surface Figure Errors} }
\begin{figure}[htbp]
    \centering
    \includegraphics[width=0.6\linewidth]{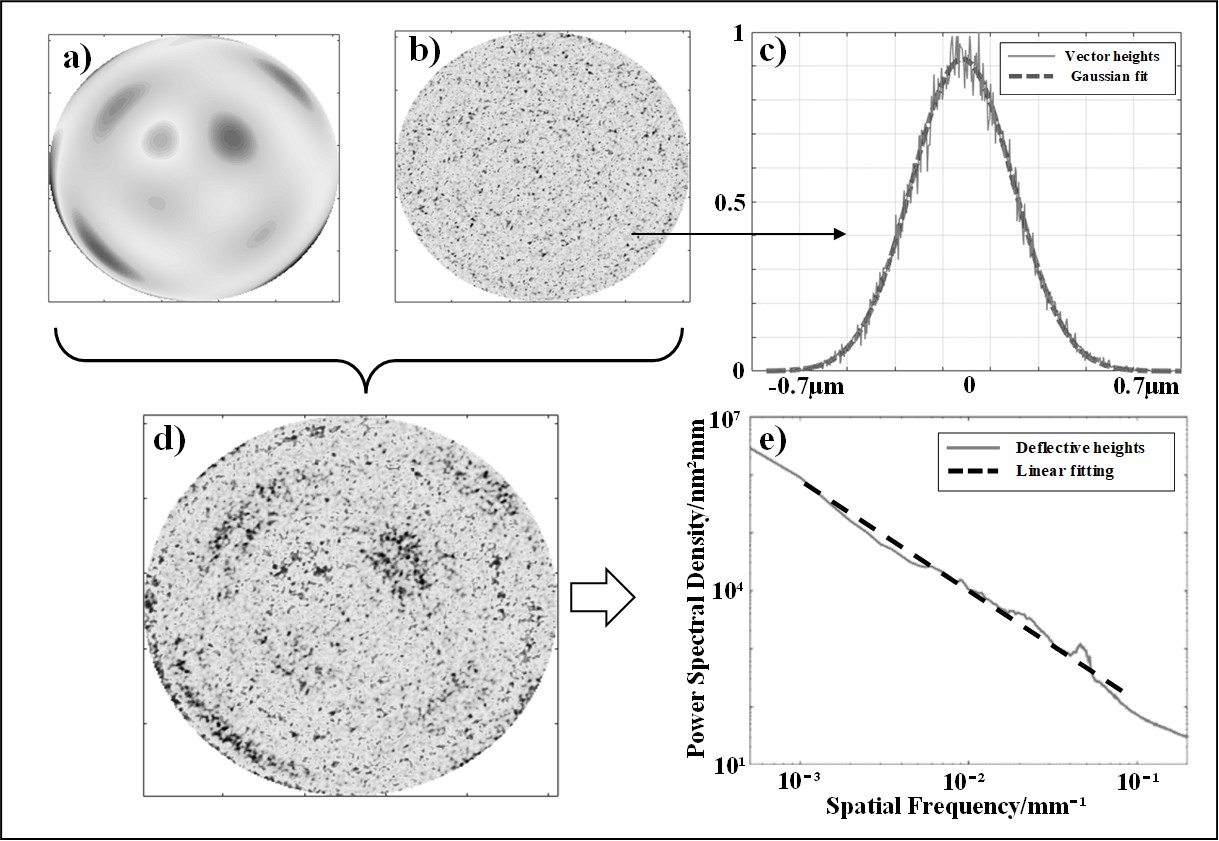}
    \caption{Full-frequency surface shape simulation process. a) represents the low-frequency 36-term Zernike fitted surface figure, b) represents the mid-to-high frequency simulated surface figure, c) represents the sagittal height distribution curve of the mid-to-high frequency surface figure, d) represents the full-frequency simulated surface figure, obtained by superposition the surface figures described in a) and b), e) represents the power spectral density (PSD) curve of the full-frequency simulated surface figure as a function of spatial frequency.}
    \label{Fig3}
\end{figure}

After processing, the surface shape of the mirror partially deviates from the intended design, leading to a degradation of the image quality of the optical system. Based on previous processing experience\citep{2010J.M.Tamkin}, mirror surface figure errors can be classified by frequency into low-frequency and medium-to-high frequency components, as illustrated in Fig.\ref{Fig3}a and Fig.3b. Low-frequency errors primarily degrade the image quality of the optical system, while the high-frequency components in the medium-to-high-frequency range diminish imaging contrast. Statistical analysis of similar historical data indicates that, after eliminating the low-frequency errors corresponding to the first 36 Zernike terms, the residual error distribution approximates a Gaussian shape, as shown in Fig.\ref{Fig3}c.

In this paper, a model was established based on the frequency characteristics of surface figure errors. First, the low-frequency errors of the mirror surface were simulated using the Zernike polynomial fitting method,while the medium-to-high-frequency errors were generated using a function-based simulation approach. Subsequently, the low-frequency and medium-to-high-frequency surface figure errors were superimposed to form the overall mirror surface figure error, as depicted in Fig.\ref{Fig3}d. The detailed calculation process is as follows: A normal distribution random matrix q was generated, with an expected value of 0 and a standard deviation of 1. By applying a filtering function to this matrix, the distribution function of the medium-to-high-frequency surface figure errors is obtained, as expressed in Eq.\eqref{eq1} below.

\begin{equation}\mathcal{\mathit{D}}\left(x,y\right)=\mathcal{\mathit{p}}\left(x,y\right)\otimes \mathcal{\mathit{q}}\left(x,y\right)\label{eq1}
\end{equation}
Here, p(x, y) represents the filtering function which can be calculated from the autocorrelation function (Eq.\eqref{eq2}), q(x, y) refers to the random matrix, and D(x, y) denotes the resulting distribution function of medium-to-high-frequency errors. 
\begin{equation}\label{eq2}
p(x,y)=\mathcal{F}^{-1}({\mathcal{F}[A(x,y)]^{1/2}})
\end{equation}
Here, A(x, y) denotes the surface autocorrelation function, which has a well-known correlation with PSD, as given in Eq.\eqref{eq3}:
\begin{equation}\label{eq3}
A(x,y)=\iint PSF(X,Y)e^{-i (Xx+Yy)}dXdY
\end{equation}
Typically, the PSD function for medium-to-high-frequency error surfaces is calculated using a functional form that includes the standard deviation and correlation length, as follows.
\begin{equation}\label{eq4}
PSF(X,Y)=\frac{l_xl_y\sigma}{4\pi}e^{-i (X^2l_x^2+Y^2l_y^2)}
\end{equation}
Here, $\sigma$ denotes the standard deviation of the surface figure simulation data, while \textit{$l_x$} and \textit{$l_y$} represent the correlation lengths in the x and y directions, respectively. By setting these parameters and applying Eq.\eqref{eq1} - Eq.\eqref{eq4}, the medium-to-high-frequency error function can be derived. The low-frequency components are then superimposed with the medium-to-high-frequency components to yield the full-band surface figure data of the mirror. The surface figure error parameters of the four mirrors are simulated in sequence corresponding to their sizes. After the full-band error simulation surface figure is obtained, its power spectral density is calculated, which shows a decreasing trend, as shown in Fig.\ref{Fig3}e.

\begin{figure}[htbp]
    \centering
    \includegraphics[width=0.75\linewidth]{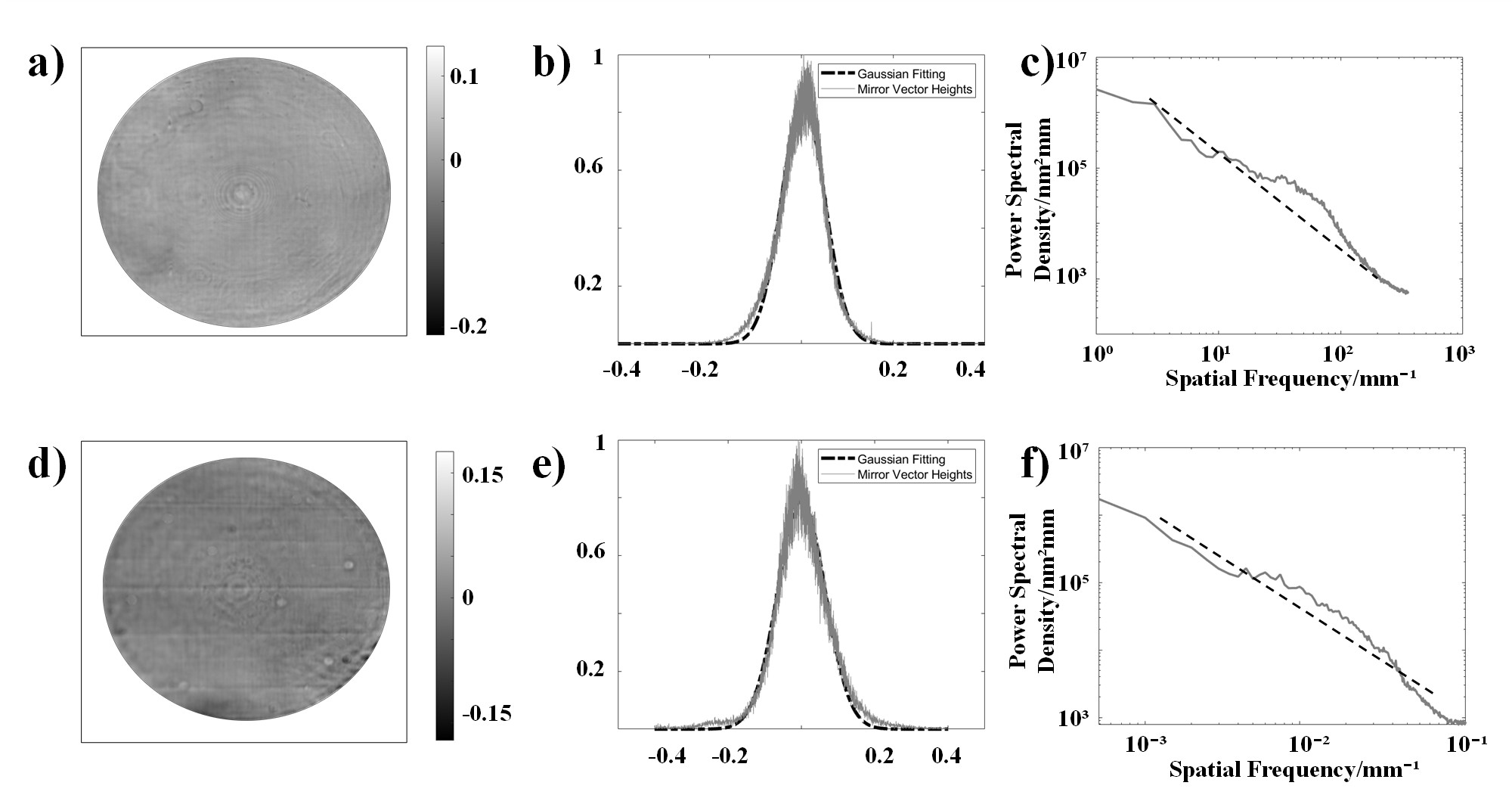}
    \caption{Analysis of frequency characteristics of mirror surface figure. The surface figure distributions are shown in Figures a) and d), the fitted curves (Figures b) and e)) both show a Gaussian distribution, the curve of PSD over the full frequency domain functions as a function of spatial frequency, as shown in Figures c) and f).}
    \label{Fig4}
\end{figure}

The frequency characteristics of the machined mirror surfaces were investigated to verify the accuracy of the aforementioned simulation method. Two large-aperture mirrors were fabricated and their surface profiles were analyzed. They were made from the same material as the primary mirror (M1) of the telescope-silicon carbide (SiC), and have circular apertures with diameters of 2 m and 1.5 m, respectively. The measured surface profiles of the mirrors are shown in Fig.\ref{Fig4}a and Fig. 4d.

First, the Zernike polynomial fitting method was employed to eliminate the first 36 low-frequency terms from both surface profile datasets. Subsequently, curve fitting was performed on the middle-to-high-frequency sagittal height components of the residual surface data. The fitted curves exhibited Gaussian distribution characteristics, as illustrated in Fig.\ref{Fig4}b and Fig. 4e, which are consistent with the simulation results in Fig.\ref{Fig3}c. Furthermore, the PSD distribution functions across the full frequency band were calculated for both surface profiles, as depicted in Fig.\ref{Fig4}c and Fig. 4f. Both curves displayed linear decreases, which are in good agreement with the simulation results in Fig.\ref{Fig3}c. These findings substantiate the validity of the mirror surface machining profile simulation modeling. By applying the surface figure error to the optical system and performing ray tracing, the induced wavefront error was calculated as 0.336$\lambda$

\subsubsection{\textbf{\textbf{Alignment Errors} } }

 Before the completion of telescope assembly, it is essential to establish an alignment and adjustment error model to assess the impact of these errors on image quality degradation. The alignment and adjustment errors consist of three components: optical element positional alignment and adjustment error, mirror installation unevenness error, and focal plane error.
 
\textit{1.Optical element positional alignment and adjustment error}

\begin{figure}[htbp]
    \centering
    \includegraphics[width=0.75\linewidth]{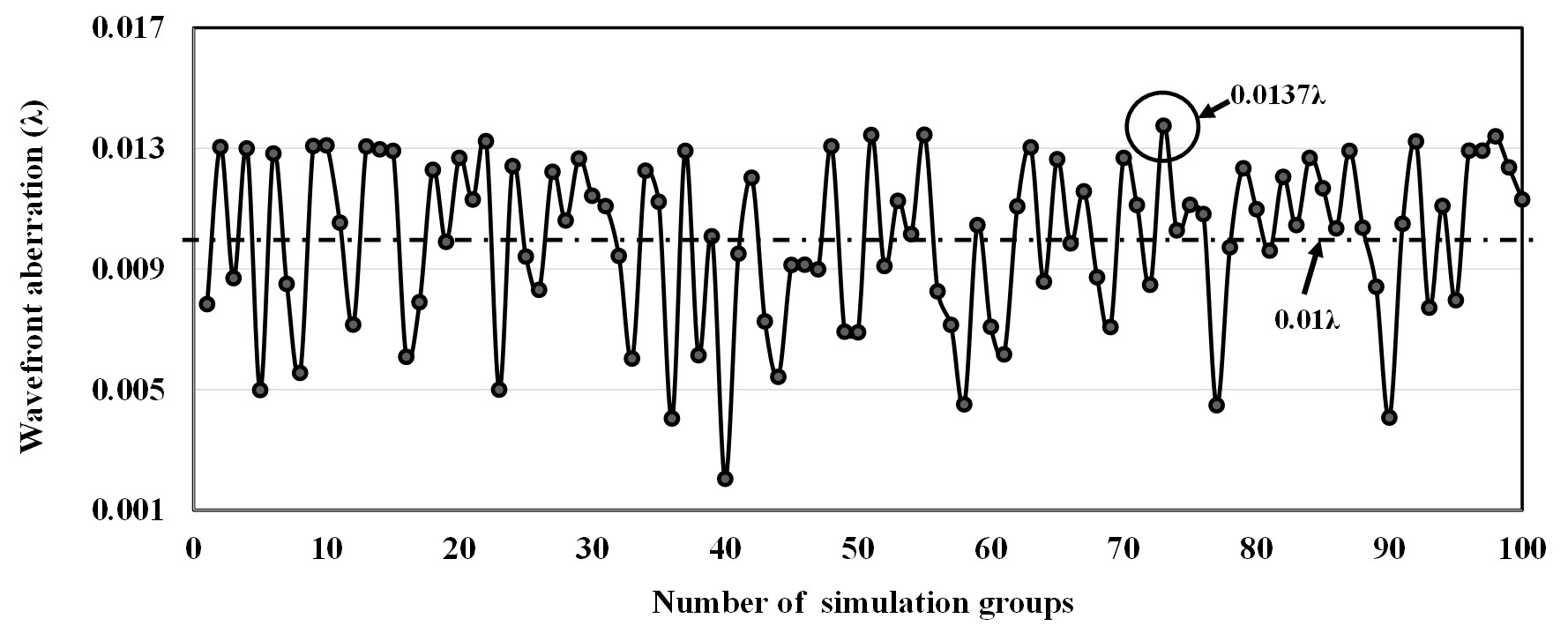}
    \caption{Statistical results of wavefront values for 100 sets of assembly and adjustment errors.}
    \label{Fig5}
\end{figure}

During the actual assembly process, discrepancies arise between the installed positions of the mirrors and their designed positions. Before the final assembly, the positional error of each mirror’s placement is not a fixed parameter but instead exists within a specified range of values. To address this, the Monte Carlo method was employed in this paper to simulate mirror installation positional errors, using uniform distribution sampling to select parameter values within the defined range intervals. The specific methodology is as follows:

Taking the primary mirror (M1) as reference, six degrees-of-freedom errors, including translations along the x, y, and z axes and rotations around the x, y, and z axes, are introduced at the installation position of the tertiary mirror (M3), thereby misaligning the optical system. Subsequently, the position parameters of the secondary mirror (M2) and the image plane are adjusted according to imaging criteria to optimize the optical system and restore imaging capability. At this stage, the position parameters of each mirror and the resulting wavefront degradation are recorded. Fig.\ref{Fig5} presents a statistical chart depicting wavefront degradation in the optical system caused by 100 sets of mirror installation position error parameters. The circular line encloses the region with the maximum wavefront value 0.0137$\lambda$, and the dashed line represents the average value 0.01$\lambda$. In this paper, the maximum value was selected to assess the extent of image quality degradation caused by this error.

\textit{2.Mirror installation unevenness error}

Taking the primary mirror (M1) as an example, the error modeling methodology was explained in detail. The primary mirror has an aperture of 2 meters, and the distance between the installation node surfaces at the rear of the mirror body is approximately 1 meter. When the installation nodes are not coplanar, it will lead to deformation of the mirror surface. As per the design requirements, non-coplanarity errors of the installation node surfaces for each mirror are assigned as 0.1 mm, 0.05 mm, 0.08 mm, and 0.05 mm, based on which the deformation parameters caused by this error are calculated. After that, the wavefront degradation resulting from this error determined as 0.024$\lambda$ based on calculation.

\textit{3.Focal plane error}

\begin{figure}[htbp]
    \centering
    \includegraphics[width=1\linewidth]{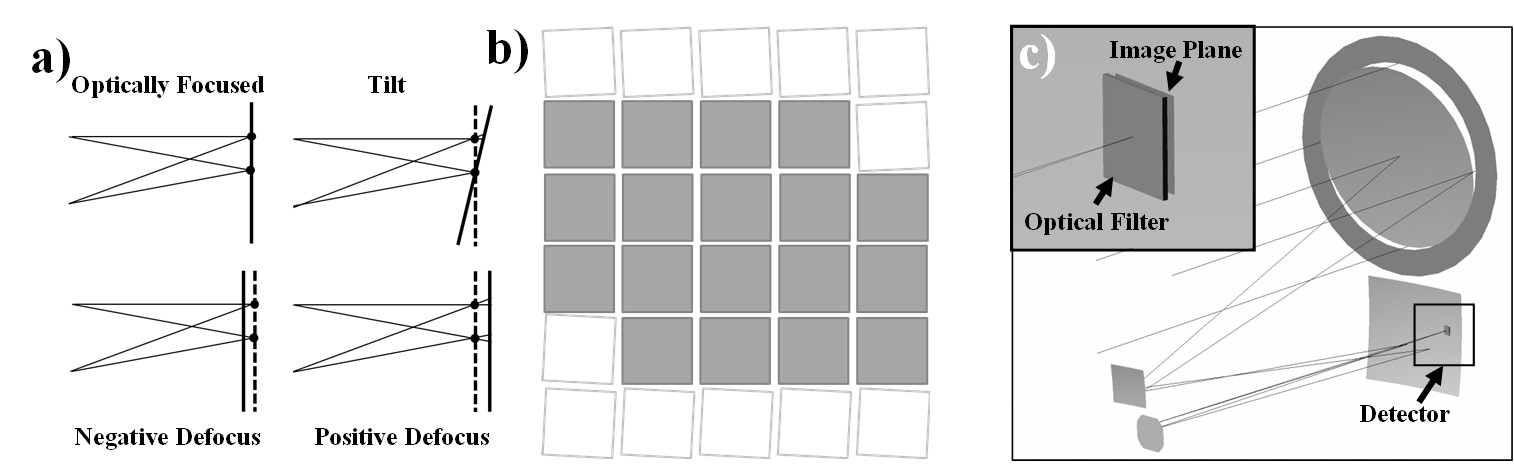}
    \caption{a) Schematic diagram of three types of focal plane errors; b) Assembly schematic of 30 CCD detector unit; c) Schematic of a CCD detector unit with a front-end filter device and its imaging optical system. The image in the upper left corner shows an enlarged partial view of the CCD detector and filter.}
    \label{Fig6}
\end{figure}

This paper first investigated the in-orbit imaging behavior of the telescope’s optical system through simulation. To comprehensively quantify the optical image quality, this paper incorporated a partial error analysis of the primary survey detectors, considering their critical role in the optical system’s performance.The primary survey detectors of the telescope consist of 30 co-focal detector units assembled in a mosaic configuration. The Peak-to-Valley (PV) simulation range for the surface unevenness of the detectors is set to ±30$\mu$m. The error sources arising from the mosaic assembly process are categorized into (1) tilt of individual detector units and (2) defocus of individual detector units. Schematic diagrams illustrating the image spot shifts induced by the detector surface under ideal focal plane alignment, tilt, and defocus conditions are presented in Fig.\ref{Fig6}a. Additionally, a filter assembly was installed in front of each detector unit, as depicted in Fig.\ref{Fig6}b. To assess the image quality of these detectors, the surface figure error of the filters must also be taken into account. In the optical system simulation model, the filter structure is implemented, and its surface figure error is incorporated. Simultaneously, by adjusting the positional parameters of the focal plane, the image quality degradation induced by detector tilt and defocus errors is quantified. Fig.\ref{Fig6}c provides a simulation schematic of a single detector unit along with its corresponding filter assembly.

\subsubsection{\textbf{\textbf{In-orbit gravity release residuals}} }
 Due to the 1 g gravitational difference between the pre-launch and post-orbit environments of a space telescope, we installed optical components in a -1 g configuration prior to launch to ensure that the telescope can image properly after entering orbit. However, the limited contact points between each optical component and the truss structure make it challenging to position all nodes of the optical components in the ideal position. To address this, this paper developed a gravity unloading residual model for orbit entry to quantify the degradation of optical image quality caused by this error. First, a finite element analysis (FEA) model was employed to define the positional parameters of each structure under a -1 g gravitational load. Subsequently, a 1 g force was applied, and the resulting structural analysis was adopted to calculate the relative positional parameters and mirror surface deformation of the optical components. Finally, these parameters were integrated into the optical system simulation model, where the wavefront degradation caused by this error was calculated to be 0.0253$\lambda$.

\subsubsection{\textbf{\textbf{In-orbit environmental change errors}} }
Due to solar irradiation, the environmental temperature of a space telescope undergoes drastic fluctuations during on-orbit operation. While the internal components of the telescope are protected by the thermal control system, minimizing temperature variations, its external structures and support platform experience significant temperature fluctuations\citep{2019T.Dauser,2003C.Kunt}. These fluctuations degrade the relative positions of optical elements and alter their surface figure. Notably, transient thermal changes constitute a form of dynamic error. In this paper, we simulated the operational environment of the telescope during on-orbit operation and evaluated the variation in its internal temperature in response to environmental changes. No obvious fluctuations were observed. Therefore, this error was treated as a static error for calculation purposes.

This paper specifically focused on calculations under extreme high-temperature conditions, particularly under scenarios with the maximum temperature variation. First, a finite element thermal model was established to calculate the temperature field distribution of the telescope under extreme high-temperature conditions during on-orbit operation\citep{2019Yang}. Subsequently, based on the calculated temperature field distribution, the relative positional parameters and mirror surface deformation parameters of the optical components were derived. These parameters were then imported into the optical system simulation model, where the wavefront degradation caused by this error was calculated to be 0.023$\lambda$.

\begin{figure}[htbp]
    \centering
    \includegraphics[width=1\linewidth]{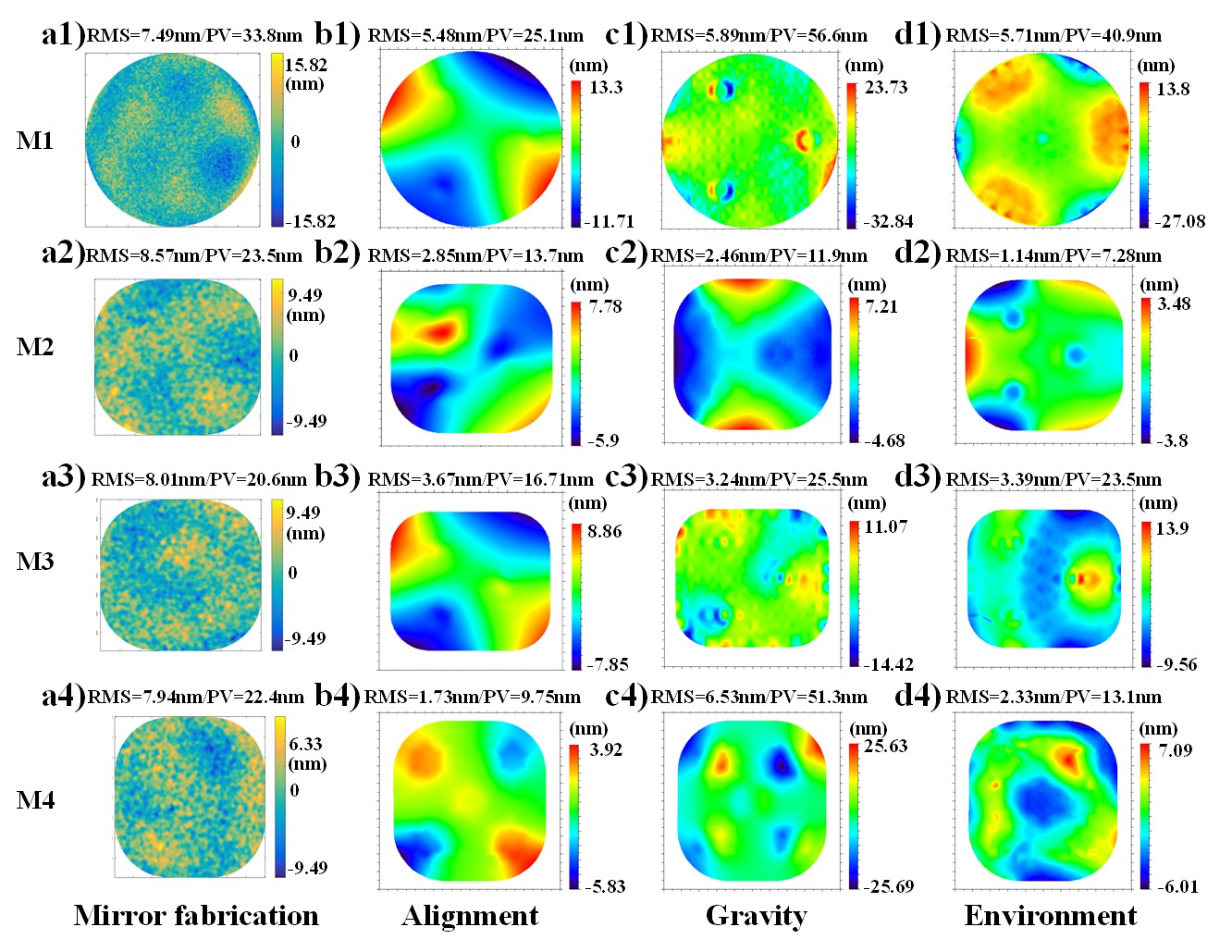}
    \caption{Schematic diagram of the mirror deformation calculated by the static error model. The deformation caused by mirror fabrication errors (Figs. a1)–a4)); the deformation caused by mirror installation tilt errors (Figs. b1)–b4)); the deformation caused by gravity unloading residuals after orbit entry (Figs. c1)–c4)); and the deformation caused by in-orbit temperature variation errors (Figs. d1)–d4)).}
    \label{Fig7}
\end{figure}

Fig.\ref{Fig7} displays the mirror deformation distribution as calculated from the static error model, including the deformation caused by mirror fabrication errors (Figs.\ref{Fig7}a1–7a4), the deformation due to mirror installation tilt errors (Figs.\ref{Fig7}b1–7b4), the deformation caused by gravity unloading residuals upon orbit entry (Figs.\ref{Fig7}c1–7c4), and the deformation resulting from in-orbit temperature variation errors (Figs.\ref{Fig7}d1–7d4). Table 1 lists the numerical values of mirror position changes, derived from the static error model, corresponding to alignment and adjustment errors, gravity unloading residuals after orbit entry, and in-orbit temperature variation errors. Here, X, Y, and Z represent the relative translation parameters along the Cartesian coordinate axes, while RX, RY, and RZ denote the relative rotation parameters around the coordinate axes. By integrating these surface and position parameters into the optical design model, the degradation of optical image quality caused by each static error model was calculated sequentially.

\begin{table}[h]
    \centering
\caption{The 6-DOF position variation caused by the change of gravity,temperature and alignment}
\label{Tabl}
    \begin{tabular}{|c|c|c|c|c|c|c|} \hline 
         &  \multicolumn{3}{|c|}{X($\mu$m)}&  \multicolumn{3}{|c|}{RX($^{\prime\prime}$)}\\ \hline 
         Optical Elements&  Gravity&  Environment&  Alignment&  Gravity&  Environment& Alignment
\\ \hline 
         M1
&  -21.0&  1.9&  0.01&  0.62&  -0.06& 0.01
\\ \hline 
         M2
&  -82&  3.35&  -209&  -2.95&  -0.55& -18.5
\\ \hline 
         M3
&  -14.6&  -0.47&  51.7&  0.07&  -0.03& -58.9
\\ \hline 
         M4
&  -171&  -0.5&  -0.06&  0.36&  0.03& 187
\\ \hline 
         Focal Plane
&  -26.7&  0.4&  0.03&  -0.39&  -1.23& -0.01
\\ \hline
 & \multicolumn{3}{|c|}{Y($\mu$m)}& \multicolumn{3}{|c|}{RY($^{\prime\prime}$)}\\\hline
 Optical Elements& Gravity& Environment& Alignment& Gravity& Environment&Alignment
\\\hline
 M1
& 2.4& 2.63& 0.15& 0.48& 0.1&0.01
\\\hline
 
M2
& -0.08& -1.75& -362& -1.08& -0.09&11.2
\\\hline
 M3
& 0.78& 3.46& 256& 0.2& -0.08&37.2
\\\hline
 
M4
& -22.2& -1.79& -0.28& 19.1& -0.08&-101
\\\hline
 Focal Plane
& -4.93& 0.33& 0.05& -4.31& -0.26&0.01
\\\hline
 
& \multicolumn{3}{|c|}{Z($\mu$m)}& \multicolumn{3}{|c|}{RZ($^{\prime\prime}$)}\\\hline
 Optical Elements& Gravity& Environment& Alignment& Gravity& Environment&Alignment
\\\hline
 
M1
& 0.26& -0.2& 0.06& 0.34& -0.07&-0.01
\\\hline
 M2
& -1.42& -0.15& 94.63& 7.02& -0.07&0.02
\\\hline
 M3
& 2.96& 3.09& 19.55& -1.74& 0.24&0.16
\\\hline
 M4
& 18.4& 2.2& -136& 7.06& -0.91&-0.03
\\\hline
 Focal Plane
& 7.76& 1.44& 0.01& 1.84& 0.17&0.06
\\\hline
    \end{tabular}
\end{table}

\subsection{\textbf{Dynamic error model} }
The images captured by space telescopes during long-duration exposures are influenced by both static and dynamic error factors. As described in Section 2.1.5, the inner-layer temperature of the telescope remains stable, with minimal variation during in-orbit operation, rendering transient thermal-induced errors negligible. Consequently, dynamic errors in this paper were primarily attributed to vibrations inherent to the telescope itself. The telescope is equipped with an image stabilization system, which partially mitigates image quality degradation caused by vibrational errors. It has been established that the stabilization control system is capable of compensating for vibrational errors when the frequency is 0-8 Hz. Therefore, the residual errors that remained after stabilization within this band were analyzed in this paper, which were referred to as ”stabilization residuals”. For vibrational errors exceeding 8 Hz, which cannot currently be eliminated, they were defined as ”micro-vibration errors” and their impact was quantified in this paper. The following sections provide a detail analysis of these two categories of dynamic errors.

\subsubsection{\textbf{\textbf{Image stabilization residuals }} }
 The telescope system compensates for image shifts caused by vibrations by adjusting the position of the fast-steering mirror (M4) in a real-time manner, thereby mitigating dynamic errors. However, due to the finite control precision of the fast-steering mirror, stabilization residuals are introduced during the correction process.
 
This study simulated the real-time low-frequency vibration correction process (0–8 Hz) of the telescope to model the stabilization residuals. First, vibrational error parameters were applied to the structural simulation of telescope, enabling the calculation of time-dependent positional variations of each optical component. The position parameters of these components at different moments were then input into the optical system simulation software to obtain the image point offset at each time step. By simulating the movement of the fast-steering mirror to compensate for the image point offsets in the optical system, the corrected low-frequency vibration error model, namely the image stabilization residual model, was finally obtained.

\subsubsection{\textbf{\textbf{Micro-vibration errors}} }
 The magnitude of micro-vibration errors at each moment is relatively small and is hard to be corrected through compensation methods. However, the cumulative effect of these small differences over a long period cannot be ignored in high-precision astronomical calculations, necessitating appropriate modeling and analysis.

A micro-vibration error model was developed in this paper by simulating the real-time high-frequency vibration process of the telescope. First, medium-to-high-frequency error parameters were introduced into the telescope system to calculate the temporal variations in the positions of the optical components. These position data at different moments were subsequently input into the optical system simulation software to generate the micro-vibration error model.

Importantly, the PSF for static errors can be directly derived from the optical system simulation model. In contrast to the static error model, the dynamic error model requires the cumulative result of the PSF images over time. The optical system of the CSST is an incoherent system, and the dynamic PSF image is equal to the superposition of the PSF images at each moment during long exposure. Therefore, in this paper, static PSF images from different moments were superimposed to obtain the dynamic PSF images. To ensure accurate sampling, the sampling interval for dynamic image calculations should be shorter than that used for micro-vibration error calculations. Given that the maximum vibration frequency of micro-vibrations is 300 Hz, according to the sampling theorem, the sampling interval for calculating the dynamic image should be shorter than or equal to 1/(2×300 Hz)$\approx$1.67 ms. For computational convenience, the image sampling interval in this paper was set to 1 ms.

\subsection{\textbf{Sampling parameters at the detector end} }
As shown in Fig.\ref{Fig6}b), the main survey detector of the CSST consists of 30 detectors. Based on the positional parameters on the focal plane, the PSF images of the observation FOV for each detector can be calculated. To balance computational efficiency and simulation accuracy, the PSF for 30×30 uniformly distributed field points on each detector unit was first calculated, and interpolation was then applied to determine the PSF for the remaining FOV. The following sections will introduce the sampling wavelengths and parameter settings used for the images of each detector unit.

\subsubsection{\textbf{\textbf{\textbf{Sampling Wavelength of the PSF} } } }
We divided the effective spectral range of the incident light for each detector unit into four equal intervals and used the central wavelength of each interval as the sampling wavelength. The PSF was then calculated for each of these sampling wavelengths. Table\ref{Tab2} presents the observation wavelength ranges for each band of the survey detector.

\begin{figure}[htbp]
    \centering
    \includegraphics[width=0.6\linewidth]{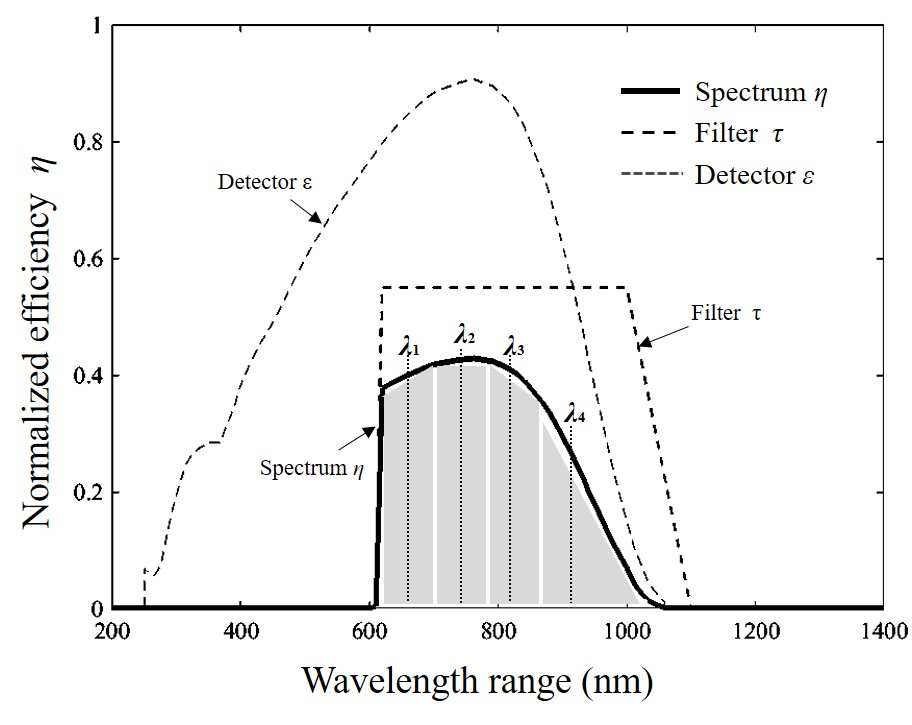}
    \caption{Spectral curve and sampling wavelength distribution of i-band detector. In the diagram, the top dashed line, the middle dashed line, and the bottom solid line denote the detector efficiency \textit{$\epsilon$}, filter transmittance \textit{$\tau$}, and the calculated spectral efficiency \textit{$\eta$}, respectively.}
    \label{Fig8}
\end{figure}

\begin{table}[h]
    \centering
     
\caption{optical wavelength range for the primary detector}
\label{Tab2}
        \begin{tabular}{|c|c|c|c|c|c|} \hline 
         No.&  Types& Wavelength range
 & No.& Types&Wavelength range
\\ \hline 
         1&  NUV& 0.255$\mu$m$\sim$0.32$\mu$m &  5& i&0.69$\mu$m$\sim$0.82$\mu$m 
\\ \hline 
         2&  u& 0.32$\mu$m$\sim$0.4$\mu$m & 6& z&0.82$\mu$m$\sim$1.0$\mu$m 
\\ \hline 
         3&  g& 0.4$\mu$m$\sim$0.55$\mu$m & 7& y&0.9$\mu$m$\sim$1.1$\mu$m \\ \hline 
         4&  r& 0.55$\mu$m$\sim$0.69$\mu$m & & &\\\hline
    \end{tabular}

\end{table}
Taking the i-band detector as an example, the effective incident wavelength range corresponds to the shaded region under the detector’s spectral response curve \textit{$\eta$}, as shown in Fig.\ref{Fig8}. The physical process of converting incident photons into an electronic image is as follows: incident photons pass through the telescope’s optical system and the filter before reaching the detector, where they are used to generate an electronic image. Therefore, the value of \textit{$\eta$} is determined by three parameters: mirror reflectivity R, filter transmittance \textit{$\tau$} , and the detector’s quantum absorption efficiency \textit{$\epsilon$} . The specific relationship can be expressed by the following equation:

\begin{equation}
\eta(\lambda)=R(\lambda)\cdot\tau(\lambda)\cdot\epsilon(\lambda)\label{eq5}
\end{equation}

\subsubsection{\textbf{\textbf{PSF Sampling Image Parameters}} }
Once the incident wavelength band is determined, the sampling parameters for the PSF image must be established. Generally, the relationship between the sampling interval in the exit pupil plane and the image plane of the optical system is described by the following equation:
\begin{equation}
Nd_p=\frac{\lambda f}{d_c}\label{eq6}
\end{equation}
where, \textit{f} is the focal length of the optical system,\textit{ $\lambda$} is the incident light wavelength, \textit{d\textsubscript{p}}  is the exit pupil sampling interval, \textit{d\textsubscript{c}} is the detector image plane sampling interval, and \textit{N} is the number of sampling points. Additionally, the exit pupil sampling interval is related to the exit pupil diameter, as expressed by:
\begin{equation}\label{eq7}
D=\frac{N}{Q}d_p
\end{equation}
Here, \textit{D} is the exit pupil diameter, and \textit{Q} is the sampling factor. Then, the below equation can be deduced:
\begin{equation}\label{eq8}
Q=\frac{\lambda f}{Dd_c}=\frac{\lambda F}{d_c}
\end{equation}
Here, \textit{f}/\textit{D} is the F-number of the optical system. According to the Nyquist sampling theorem, the sampling frequency must be at least twice the signal frequency, i.e., \textit{$Q\geq2$}  to fully preserve the information of the original signal. Therefore, the detector sampling interval should satisfy the following equation:
\begin{equation}\label{eq9}
d_c\leq\frac{\lambda F}{2}
\end{equation}
Based on optical design principles and experience, it is understood that reducing the image sampling interval can lead to a decrease in the system’s Modulation Transfer Function (MTF) value and imaging quality. According to Eq.\eqref{eq8} and Eq.\eqref{eq9} in the text, the shorter the wavelength of the incident light, the smaller the maximum allowable sampling interval dc. To simultaneously satisfy the sampling interval requirements across different optical bands, the minimum ultraviolet wavelength should be selected as the critical value. The minimum sampling interval corresponding to the ultraviolet band with a wavelength of 0.255$\mu$m is approximately 1.785$\mu$m. At the initial stage of the simulation, the sampling intervals for the NUV and u-band detectors were set to 1.25$\mu$m, while the intervals for other detectors were configured at 2.5$\mu$m. In practical detector designs, where the pixel size is 10$\mu$m, a 2.5$\mu$m sampling interval achieves a 4× oversampling ratio. Furthermore, when comparing the ultraviolet band with sampling intervals of 1.25$\mu$m and 2.5$\mu$m, the differences in PSF ellipticity and REE80 values were found to be minimal. However, both configurations resulted in a 4× increase in data storage requirements and computation time. Therefore, the sampling interval of the PSF images was uniformly assigned to 2.5$\mu$m across all bands, with a fixed sampling grid size of 256×256.
 
\subsubsection{\textbf{\textbf{Ellipticity calculation} } }
Ellipticity is an important physical parameter in astronomical calculations\citep{1997C.R.Keeton}. In this paper, the ellipticity of the PSF was calculated by centering the PSF on its centroid and considering a radius of 0.5 arcseconds, with the specific formula as follows. It is important to emphasize that 0.5 arcseconds is a predefined parameter. While this parameter could be adjusted to 1, 1.5, or other reasonable values, the resulting REE80 values would vary depending on the selected parameter. To ensure consistency across simulations, all data processing teams adopted this standardized parameter prior to simulations. First, the centroid position of the PSF was calculated:
\begin{equation}\label{eq10}
\left\{
\begin{aligned}
\overline{x} &=\frac{\sum\limits_{x,y}{x}P{(x,y)}W{(x,y)}}{\sum\limits_{x,y}{P{(x,y)}W{(x,y)}}}\\\\
\overline{y} &=\frac{\sum\limits_{x,y}{y}P{(x,y)}W{(x,y)}}{\sum\limits_{x,y}{P{(x,y)}W{(x,y)}}}
\end{aligned}
\right.
\end{equation}

Here, \textit{P}$(x, y)$ represents the light intensity of the PSF at the focal plane position $(x, y)$, and \textit{W}$(x, y)$ is the weight function. From this, the second-order moments of the PSF can be obtained:

\begin{equation}\label{eq11}
\left\{
\begin{aligned}
Q_{xx} &=\frac{\sum\limits_{x,y}{(x-\overline{x})^2}P{(x,y)}W{(x,y)}}{\sum\limits_{x,y}{P{(x,y)}W{(x,y)}}} &\\\\
Q_{yy} &=\frac{\sum\limits_{x,y}{(y-\overline{y})^2}P{(x,y)}W{(x,y)}}{\sum\limits_{x,y}{P{(x,y)}W{(x,y)}}} &\\\\
Q_{xy} &=\frac{\sum\limits_{x,y}{(x-\overline{x})}{(y-\overline{y})}P{(x,y)}W{(x,y)}}{\sum\limits_{x,y}{P{(x,y)}W{(x,y)}}}
\end{aligned}
\right.
\end{equation}

The two components of the PSF ellipticity were calculated using the following equation:
\begin{equation}\label{eq12}
\left\{
\begin{aligned}
e_1 &= \frac{Q_{xx}-Q_{yy}}{Q_{xx}+Q_{yy}} &\\
e_2 &= \frac{2Q_{xy}}{Q_{xx}+Q_{yy}}
\end{aligned}
\right.
\end{equation}
The PSF ellipticity value was then calculated\citep{2014Zeng}:
\begin{equation}\label{eq13}
e=\sqrt{e_1^2+e_2^2}
\end{equation}

\section{Image quality analyses}

\begin{figure}[htbp]
    \centering
    \includegraphics[width=0.75\linewidth]{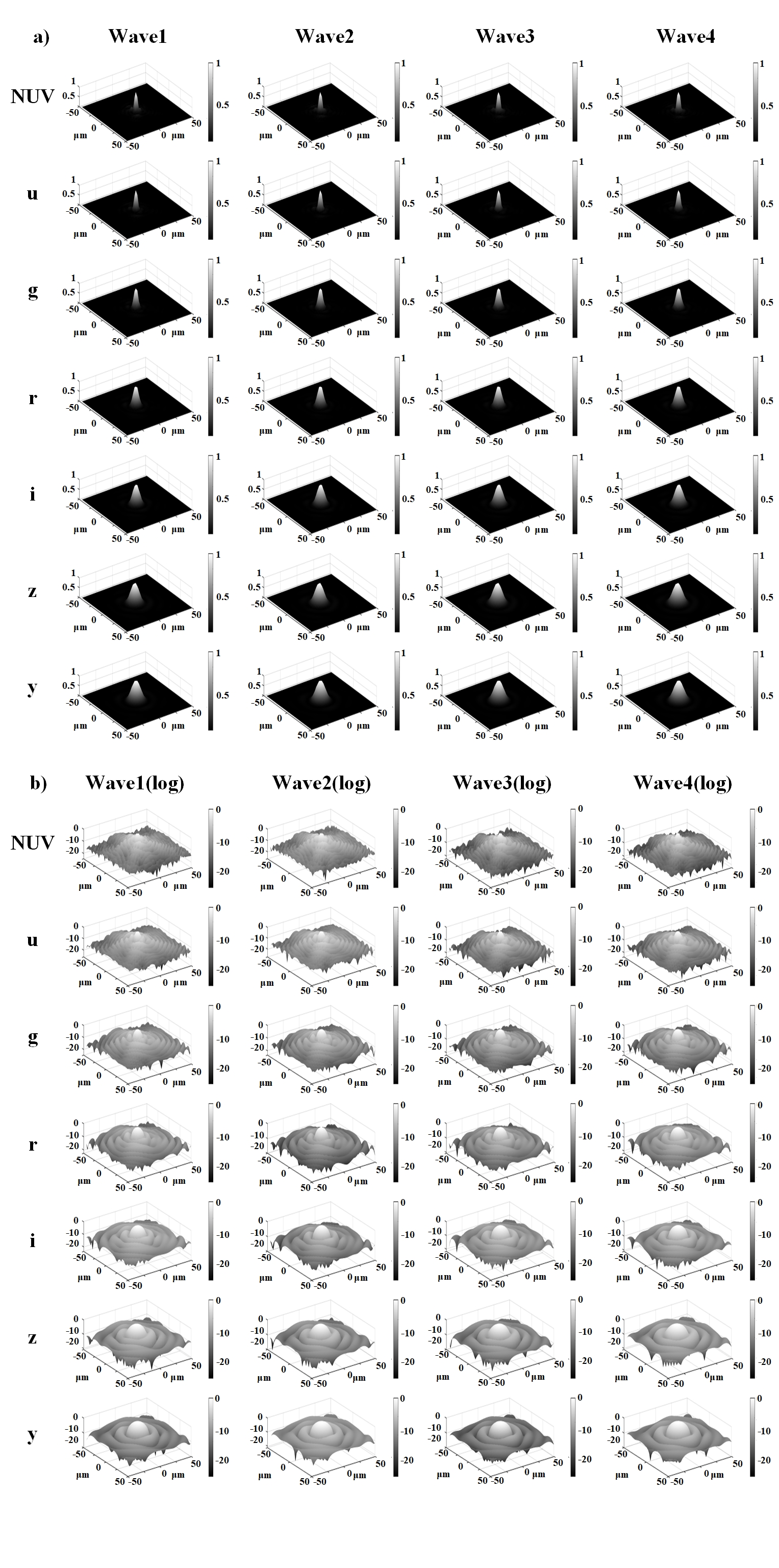}
    \caption{a) Shows the PSF generated by the four sampled wavelengths in the NUV, u, g, r, i, z and y bands. The units of the x- and y- axis are $\mu$m, and the z-axis represents the normalized intensity values. b) Shows the PSF corresponding to a) plotted on a natural logarithmic scale.}
    \label{Fig9}
\end{figure}

After the end-to-end simulation modeling of the telescope optical system, we analyzed its image quality. Specifically, five indicators were calculated to evaluate the optical image quality of the space telescope. These indicators included static and dynamic PSF, the ellipticity distribution of the PSF over the entire FOV, the wavefront degradation amounts of various error models and field distortion over the entire FOV. The following analysis proceeds sequentially through three aspects: static image quality, dynamic image quality, and image field distortion.

\subsection{\textbf{\textbf{Static image quality} } }

First, the PSF of the telescope was computed for different sampling wavelengths within the same observation FOV, as shown in Fig.\ref{Fig9}. The wavelengths corresponding to wave 1 to wave 4 correspond to the four central wavelength values specified in Section 2.3.1, with these values increasingly progressively. According to the parameters in Table \ref{Tab2}, the wavelengths span the NUV, u, g, r, i, z, and y bands, in that order. The calculation results reveal that within the same band, variations in the PSF spot are relatively small. However, when the observation band is changed, a significant change in the PSF spot shape is observed. Meanwhile, the above simulation results are consistent with the theoretical calculation results described in Eq.\eqref{eq6} - Eq.\eqref{eq8}, where, under the same sampling parameters, the size of the PSF spot increases gradually with the wavelength.

\begin{figure}[htbp]
    \centering
    \includegraphics[width=0.75\linewidth]{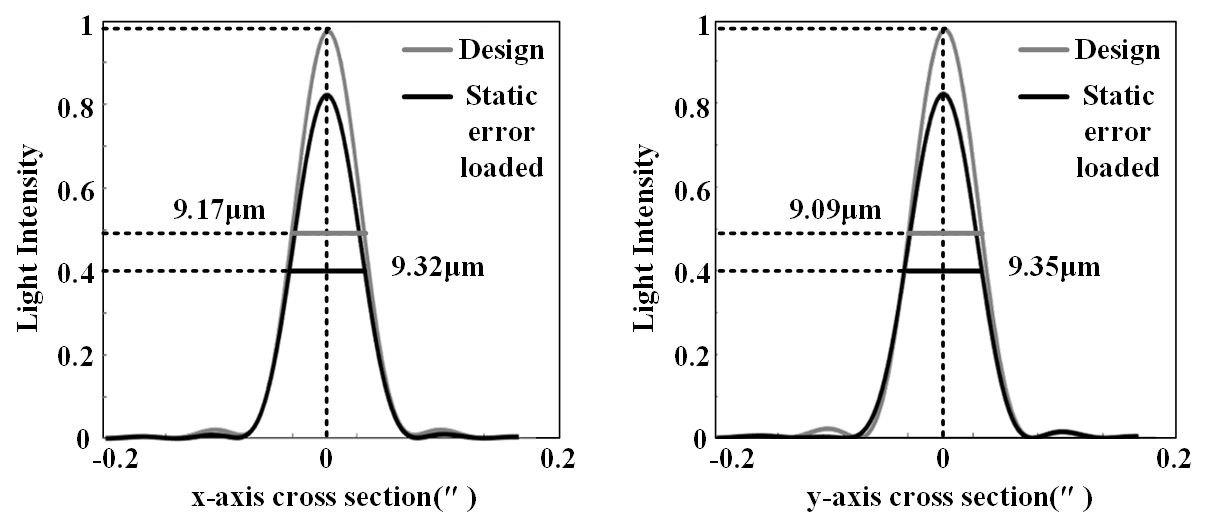}
    \caption{Comparison of cross-sectional profiles along the x-axis and y-axis for the center of FOV PSF of the telescope under design and static error conditions, with the incident light set to monochromatic radiation at 632.8 nm}
    \label{Fig10}
\end{figure}

In this section, the changes in the telescope’s PSF under the design and static error model conditions were compared. The PSF at the center of the FOV was calculated, with the incident light sampling wavelength set to 632.8 nm. Fig.\ref{Fig10} presents the cross-sectional profiles of the PSF at the center of the FOV along both the x-axis and y-axis. The Strehl ratios of the PSF under the design and static error conditions are 0.964 and 0.816, respectively, indicating that the PSF under the static error condition exhibits energy dispersion compared to the initial PSF, with a corresponding increase in the full width at half maximum (FWHM) value. This suggests that static errors have compromised the image quality of the optical system, reducing the energy enclosed within the telescope’s PSF.

\begin{figure}[htbp]
    \centering
    \includegraphics[width=1\linewidth]{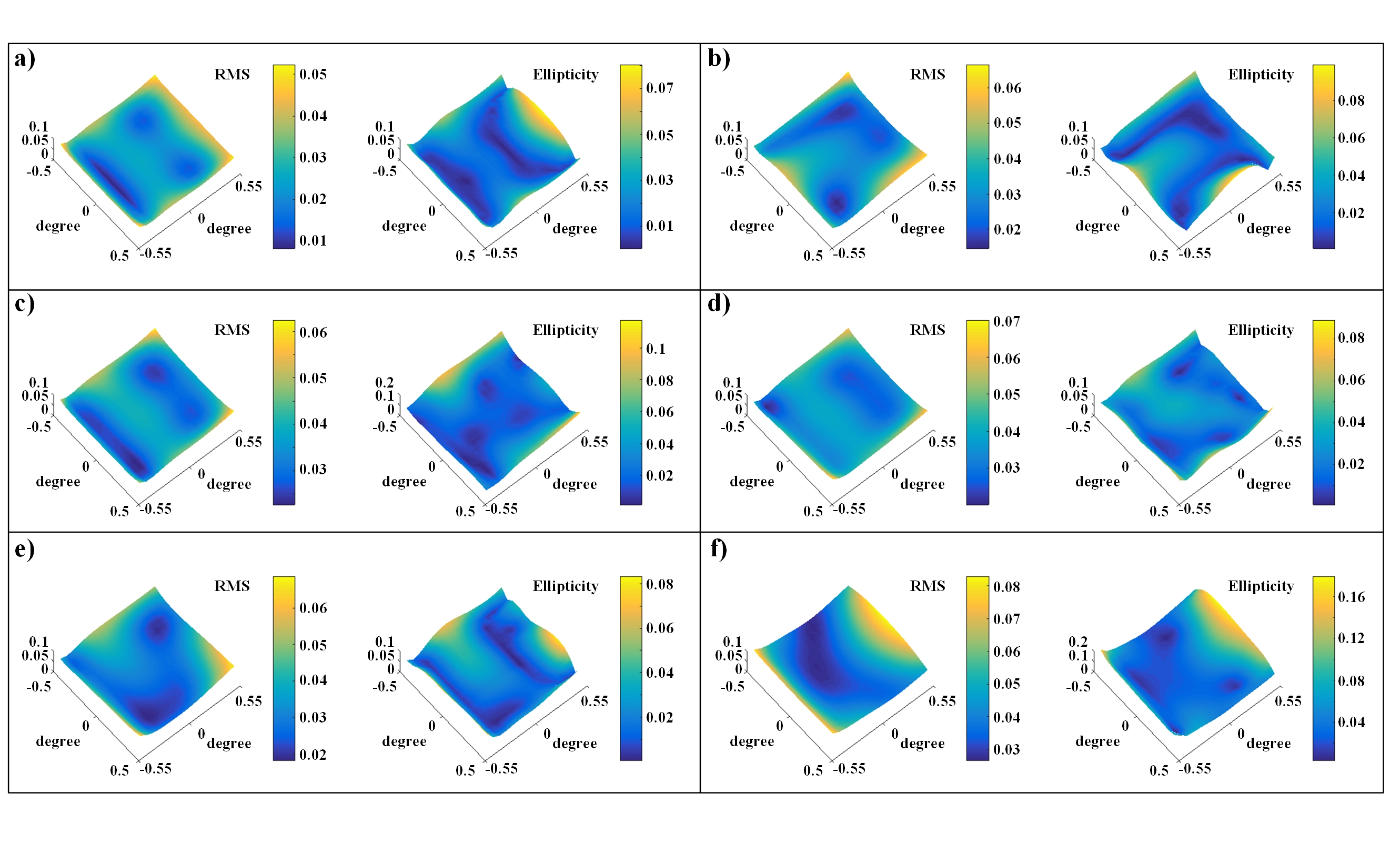}
    \caption{Comparison of the distributions of the RMS values of the wavefront and the ellipticity values across the full FOV under different error conditions. a) Optical system design, b) Mirror surface sharp errors, c) In-orbit variable environment, d) In-orbit variable gravity, e) Mirror installation unevenness, f) Alignment error}
    \label{Fig11}
\end{figure}

The ellipticity of the PSF has a known computational relationship with the wavefront value of the optical system \citep{2014Zeng}. To this end, this paper calculated the ellipticity distribution of the PSF across the full FOV and compared the distribution of ellipticity values of the PSF with the wavefront error values across the full FOV. The results are detailed in Fig.\ref{Fig11}. The ellipticity values and wavefront error values of the PSF across the full FOV were calculated for five static error models. Among them, the alignment and adjustment errors were divided into mirror position errors and mirror installation tilt errors for calculation. Based on the calculation results, we observed that the distributions of the ellipticity values of the PSF and the wavefront error values are similar. By adjusting the wavefront distribution across the full FOV, the distribution of the ellipticity values of the PSF across the full FOV can be indirectly controlled, allowing for the adjustment of ellipticity uniformity to meet the development requirements of the instrument.

In this section, the degradation of optical image quality caused by different static error models, as well as the degradation of REE80 angle radius of the PSF, was calculated. The calculation results, outlined in Table \ref{Tab3}, reveal that the largest wavefront degradation and the REE80 angle radius degradation are caused by mirror surface figure errors . The wavefront degradation caused by different factors, including gravity, temperature, and installation tilt errors, shows similar magnitude, while that caused by alignment and adjustment errors is the smallest. Therefore, when the optical performance of the telescope system described in this paper did not meet the development requirements, the priority should be given to improving the mirror manufacturing process to enhance the imaging performance of the telescope system.

\begin{table}[h]
    \centering
    \caption{Comparison of wavefront aberration and REE80 degradation of PSF caused by static errors.}
\label{Tab3}
\begin{tabular}{|c|c|c|} \hline 
         Static errors&  Wavefront Degradation& REE80 Degradation
\\ \hline 
         Optical design system&  better than  $\lambda $/30& better than $0.1^{\prime\prime}$\\ \hline 
         Mirror Surface Figure Errors&  0.0336 $\lambda $& $0.0229^{\prime\prime}$\\ \hline 
         Alignment and adjustment errors&  0.0137 $\lambda $& $0.0044^{\prime\prime}$\\ \hline 
         Mirror installation unevenness error&  0.0240 $\lambda $& $0.0035^{\prime\prime}$\\ \hline 
         In-orbit gravity release residuals&  0.0253 $\lambda $& $0.0095^{\prime\prime}$\\ \hline 
         In-orbit environmental change errors&  0.0230 $\lambda $& $0.0179^{\prime\prime}$\\ \hline
    \end{tabular}
\end{table}
\subsection{\textbf{\textbf{Dynamic image quality} } }

The dynamic and static error terms can be combined to comprehensively evaluate the image quality of the telescope. The degradation of image quality caused by each dynamic error model was analyzed, and the dynamic errors described in this paper were divided into image stabilization residuals and micro-vibration errors. The degradation of image quality caused by image stabilization residuals is addressed in previous work \citep{2025Li}, and the degradation caused by micro-vibration errors was calculated in this section.

\begin{figure}[htbp]
    \centering
    \includegraphics[width=0.9 \linewidth]{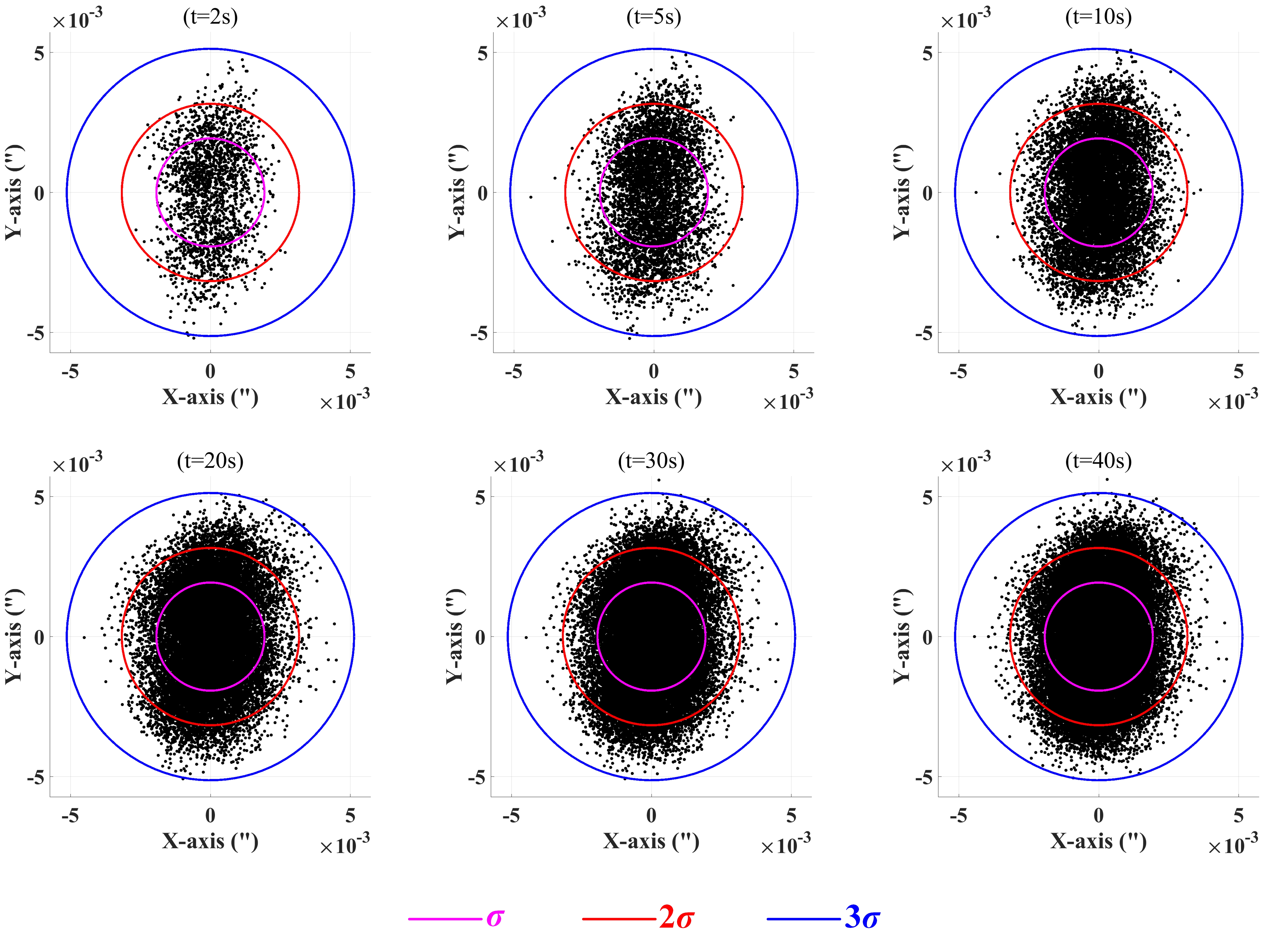}
    \caption{The calculation of the LOS jitter in the central field of view at different times. The area within the purple circular line represents the 1$\sigma$ statistical interval, the area within the red circular line denotes the 2$\sigma$ statistical interval, and the area within the blue circular line corresponds to the 3$\sigma$ statistical interval.}
    \label{Fig12}
\end{figure}

Typically, telescope exposure times last several minutes. Fig.\ref{Fig12} shows a schematic diagram of the change in the line-of-sight (LOS) jitter at the center of the FOV caused by micro-vibration errors at different moments. The figure indicates that after an exposure duration of more than 20 seconds, there is no significant change in the region where the spot moves. Statistical analysis of the 3$\sigma$ values of the LOS jitter for six different exposure durations shows that the values are all lower than 0.01 arcseconds. This suggests that after the PSF spot stabilizes, the jitter's influence on image quality is minimal, allowing the selection of the shortest exposure duration for further evaluations. By focusing on this duration, the efficiency of simulation calculations can be improved.

\begin{figure}[htbp]
    \centering
    \includegraphics[width=1\linewidth]{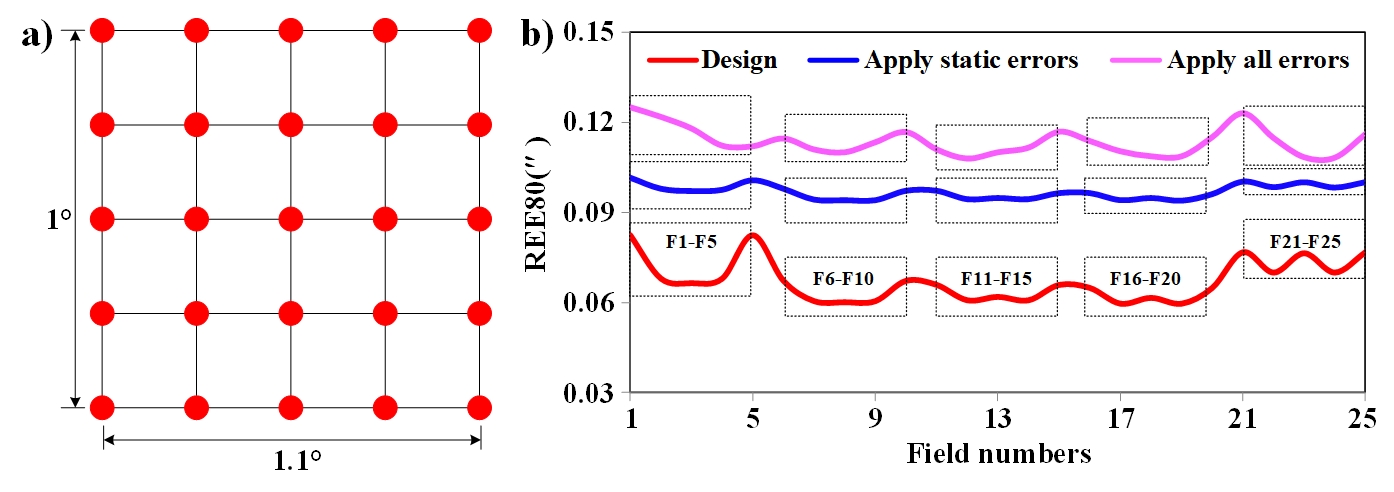}
\caption{Left: sampling field of view position. Right: calculated REE80 values for 25 FOV under different states. The dark gray curve at the bottom corresponds to the initial design state, the light gray curve in the middle represents the static error state, and the black curve at the top indicates the full error state.}
\label{Fig13}
\end{figure}

 The REE80 angular radius of the PSF is a key indicator for evaluating the imaging performance of a telescope. Fig.\ref{Fig13}a) the process of calculating the REE80 values across 25 field points in the full FOV under static and dynamic error conditions. The telescope in this paper uses an off-axis three-mirror optical system. In its initial state, the PSFs at symmetric field points (e.g., F1 and F5) are similar in appearance, and their REE80 values are also the same. As shown in Fig.\ref{Fig13}b), the lowest curve represents the symmetry of segments within the dashed-line boxes. However, as static and dynamic error models are applied, the symmetry of these segments is reduced. This is because the superposition of various model errors causes the wavefront distribution of the optical system to degrade in a non-symmetric manner, which hinders high-order optimization of the system. The computational results reveal that under the design condition, the average REE80 value across 25 field measures 0.067 arcseconds. When static errors are applied, the average REE80 value degrades to 0.097 arcseconds, representing +0.03 arcseconds increase compared to design condition. When dynamic errors are further introduced, the REE80 value deteriorates to 0.114 arcseconds, representing +0.046 arcseconds increase compared to design condition. In addition, after applying both static and dynamic errors, the distribution of the REE80 values of the PSF across the 25 fields of the telescope remains continuous without abrupt changes.

\subsection{\textbf{\textbf{Field distortion modeling} } }
\begin{figure}[htbp]
   \centering
  \includegraphics[width=16cm, angle=0]{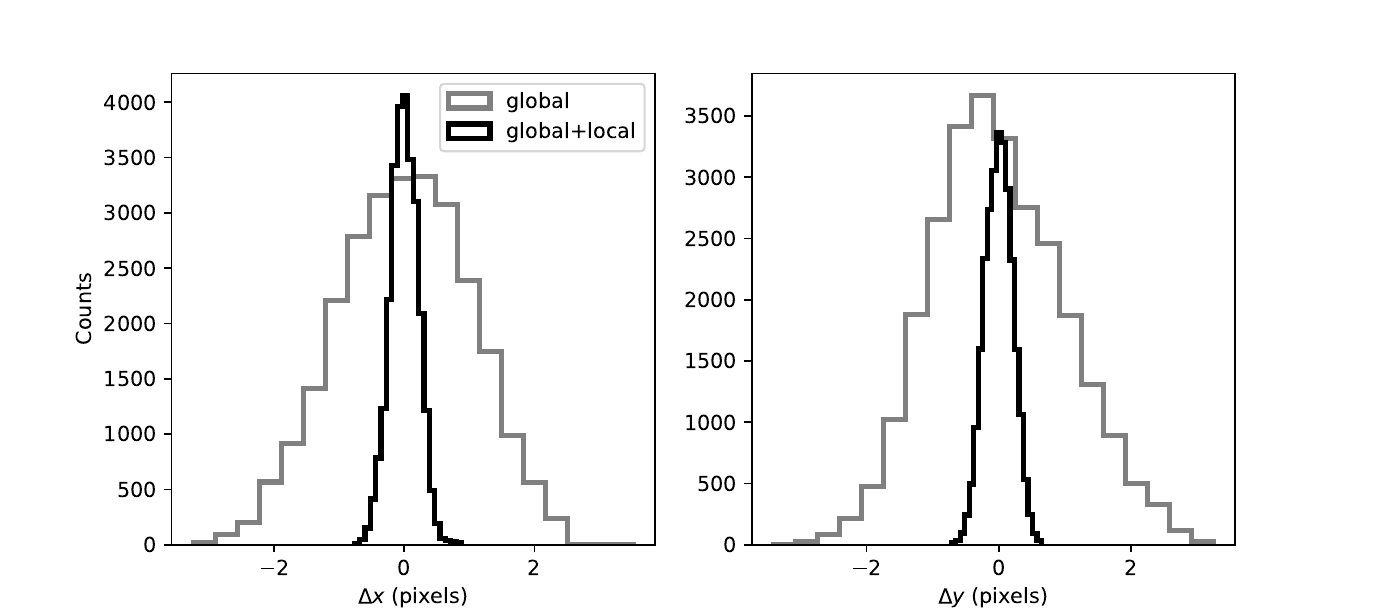}
   \caption{{Residual distributions of the field distortion model {{along the $x$-axis (left) and $y$-axis (right)}}, where the gray and black histograms represent the results of global and global+local interpolations, respectively.}} 
   \label{fig:fd_resi}
   \end{figure}

Based on the descriptions above, the PSF sampling data under the full error conditions can be obtained, including the sampling coordinates $(u, v)$ and pixel coordinates $(x, y)$ of the wavefront on the pupil and focal planes, respectively. Notably, the pixel coordinates $(x, y)$ contain the field distortion caused by the optical system as well as the static and dynamic error models described above. Consequently, they could be utilized to construct a full-view distortion model for the entire focal plane. Actually, after the celestial object is projected onto the focal plane through the World Coordinate System \citep[WCS;][]{2002A&A...395.1077C,2002A&A...395.1061G}, its pixel coordinate without considering image distortion can be represented as $(x_{0}, y_{0})$, which satisfies the Eq. (14) below:

\begin{equation}
\left( \begin{array}{cc} x_{0} \\
y_{0} \end{array} \right) = 
\frac{1}{\Delta}
\left( \begin{array}{cc} \mathrm{CD2\_1} & - \mathrm{CD1\_1} \\
- \mathrm{CD2\_2} & \mathrm{CD1\_2} \end{array} \right) 
\left( \begin{array}{cc} u \\
v \end{array} \right) + 
\left( \begin{array}{cc} \mathrm{CRPIX1} \\
\mathrm{CRPIX2} \end{array} \right),
\end{equation}
where the $\mathrm{CDi\_j}$ keywords encode the rotation and scaling in image header; $\mathrm{CRPIX1}$ and $\mathrm{CRPIX2}$ represent the origin of the pixel coordinate along the $x-$ and $y-$axes, respectively; and $\Delta$ is defined as
\begin{equation}
\Delta = \mathrm{CD1\_2}\times \mathrm{CD2\_1} - \mathrm{CD1\_1} \times \mathrm{CD2\_2}. \nonumber
\end{equation}
The field distortion model is the mathematical mapping between  $(x_{0}, y_{0})$ and $(x, y)$, satisfying the below equation:
\begin{equation}
\label{eq:fdmodel1}
x = f_{x}(x_{0}, \, y_{0}), \quad y = f_{y}(x_{0}, \, y_{0}).
\end{equation}
Notably, after applying two-dimensional spline interpolation with an order of $n=4$ to approximate $f_{x}$ and $f_{y}$ over the entire focal plane, the interpolated functions $\widetilde{f_{x}}$ and $\widetilde{f_{y}}$ could be yielded. To evaluate the interpolation accuracy, residuals $\Delta{x}=x-\widetilde{f_{x}}$ and $\Delta{y}=y-\widetilde{f_{y}}$ were calculated, with the mean values of $\langle\Delta{x}\rangle = -7.6\times 10^{-10}$ pixels and $\langle\Delta{y}\rangle = -5.2\times 10^{-10}$ pixels, and the standard deviations of $\sigma_{\Delta{x}}=1.0$ pixel and $\sigma_{\Delta{y}}= 1.0$ pixel, respectively. The residual distributions are illustrated in Fig.~\ref{fig:fd_resi} (gray histograms).

\begin{figure}[htbp]
   \centering
  \includegraphics[width=7cm, angle=0]{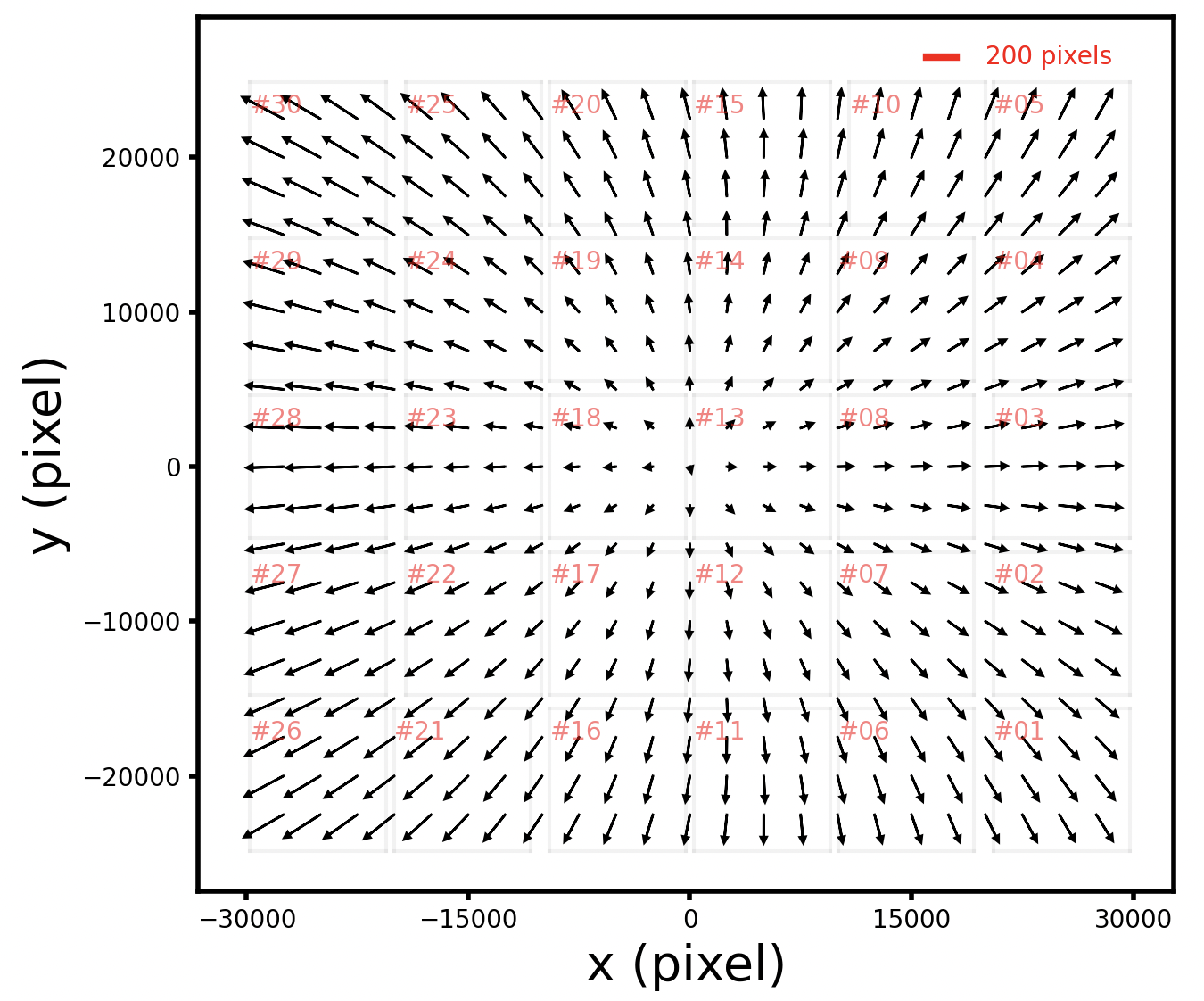}
  \includegraphics[width=7cm, angle=0]{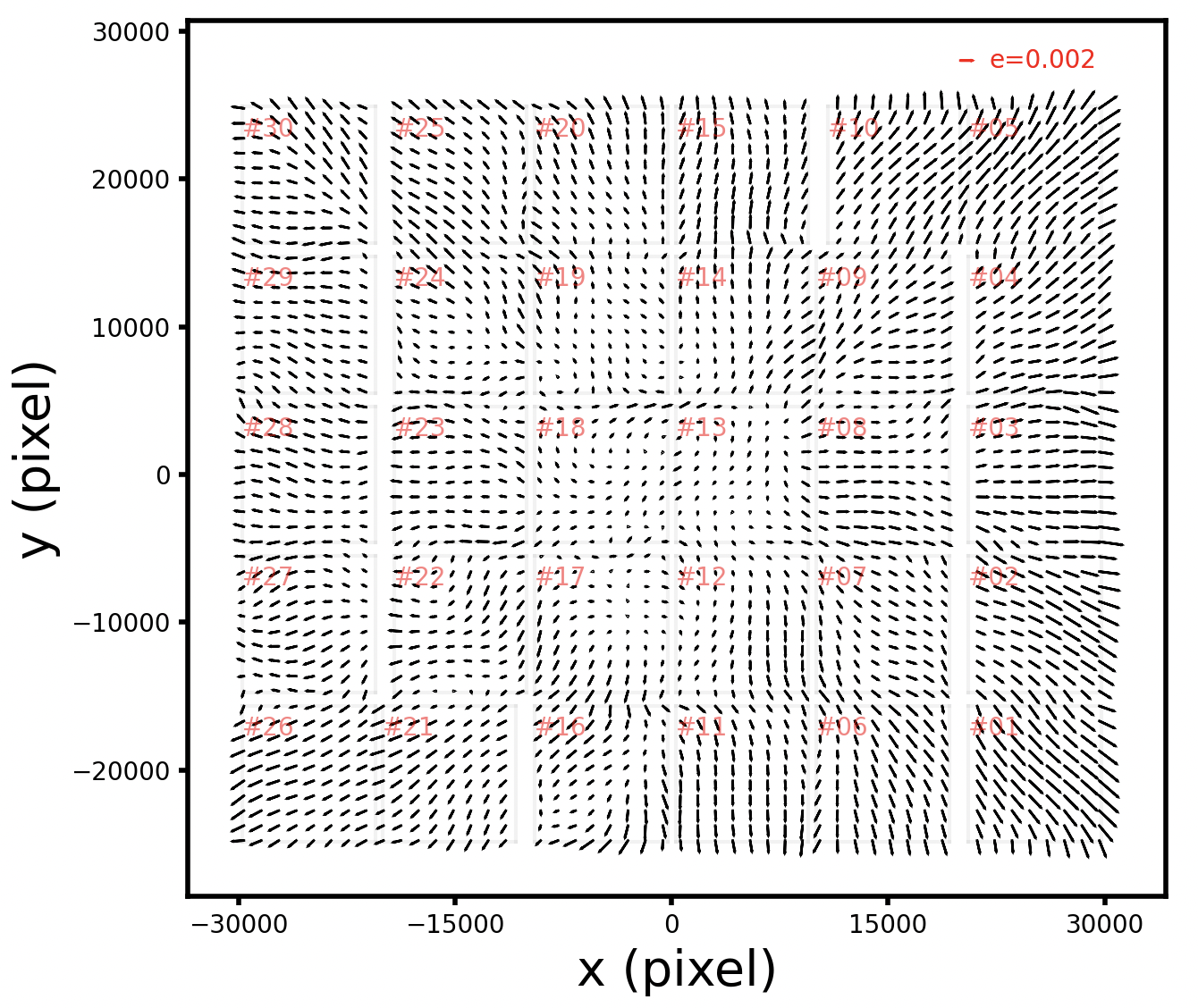}
   \caption{{Left panel: distribution of field distortion model over the entire focal plane. Right panel: distribution of ellipticty induced by field distortion over the entire focal plane. In each panel, the gray boxes are the layout of the detector units, and the number IDs are also encoded as red.}}
   \label{fig:fd_model}
   \end{figure}

Due to an unevenness error in the detector installation, the detector units in the focal plane exhibit imperfect alignment. Notably, this misalignment induces discontinuities in the field distortion between adjacent detector units, which cannot be accurately modeled by global interpolation methods. Consequently, spline interpolation was employed to further model $\Delta{x}$ and $\Delta{y}$ at the individual detector level. Then they were incorporated into Eq.\eqref{eq:fdmodel1}s as correction term. This approach is referred to as global+local interpolation. As a result, the final field distortion model is approximated as

\begin{equation}
\label{eq:fdmodel2}
f_x \simeq \widetilde{f_{x}}(x_{0}, \, y_{0}) + \widetilde{f_{\Delta{x}}}(x_{0}, \, y_{0}), \quad f_y \simeq \widetilde{f_{y}}(x_{0}, \, y_{0}) + \widetilde{f_{\Delta{y}}}(x_{0}, \, y_{0}),
\end{equation}
where $\widetilde{f_{\Delta{x}}}$ and $\widetilde{f_{\Delta{y}}}$ represent the polynomial interpolation of $\Delta{x}$ and $\Delta{y}$, respectively. As shown by the black histograms in Fig.~\ref{fig:fd_resi}, the interpolation accuracy improved significantly, with the mean residuals decreased to $\langle\Delta{x}\rangle=-3.8\times 10^{-15}$ pixels and $\langle\Delta{y}\rangle=4.0\times 10^{-15}$ pixels, respectively. The standard deviations along both axes are 0.2 pixels, representing a fivefold improvement over the global interpolation method. The left panel of Fig.~\ref{fig:fd_model} illustrates the field distortion model across the focal plane. Each arrow represents a vector, with the starting point at pixel position $(x_{0}, y_{0})$ and the end point indicates the corresponding position influenced by the field distortion. It is evident that the field distortion increases from the center to the edge of the focal plane, with the maximum position offset exceeding 200\,pixels.

Here, the ellipticity distribution induced by the field distortion was further computed using the method presented in \cite{2019ApJ...875...48Z}. Specifically, the two ellipticity components are expressed as follows:

\begin{equation}
e_{1} = +(\partial{y}/\partial{y_{0}} - \partial{x}/\partial{x_{0}})/(\partial{y}/\partial{y_{0}} + \partial{x}/\partial{x_{0}}),\nonumber
\end{equation}
\begin{equation}
\label{eq:ellmodel}
e_{2} = -(\partial{y}/\partial{x_{0}} - \partial{x}/\partial{y_{0}})/(\partial{y}/\partial{y_{0}} - \partial{x}/\partial{x_{0}}),
\end{equation}
where the symbol ``$\partial$'' represents partial derivative. Substituting Eq.\eqref{eq:fdmodel2} into Eq.\eqref{eq:ellmodel}, ellipticity at any given point in the focal plane can be determined. The right panel of Fig.~\ref{fig:fd_model} displays the ellipticity distribution across the focal plane. As with the field distortion, ellipticity also increases towards the edges. In addition, because each detector unit is assigned a unique filter, the filter difference between two adjacent detector units leads to complicated patterns and significant discontinuities in the ellipticity distribution.

\section{Conclusions}
\label{sect:summary}
In this paper, the image quality of large-aperture off-axis reflective space telescopes was simulated and analyzed under in-orbit operational conditions. Based on the types of error models, five static error models and two dynamic error models were established and integrated into a comprehensive system model. Subsequently, the PSF of the telescope and their ellipticity values, wavefront values, and field distortion were calculated under the conditions defined by these error models. Additionally, the REE80 values of the PSF for 25 uniformly distributed field points across the full FOV were calculated. The results demonstrate that after application of these errors, the PSF experiences energy dispersion, with an increase in the FWHM value. The distribution of the ellipticity values of the PSF across the full FOV shows a similar pattern to that of the wavefront values, indicating that adjustments to the wavefront distribution can indirectly influence the ellipticity distribution.As static and dynamic error models are progressively applied, the symmetry between the PSFs of the left and right fields of the telescope diminishes. The REE80 value increases by 0.046 arcseconds under the full error loaded condition. The 3$\sigma$ value of LOS jitter at the center of FOV is below 0.01 arcseconds. Furthermore, due to varying incident wavelengths across the detector units, the PSF ellipticity values exhibit weak continuity at their adjacent regions.

By employing an end-to-end simulation and analysis approach, the imaging quality of a space telescope under the influence of comprehensive error factors can be effectively evaluated. Moreover, by quantifying the wavefront degradation associated with various error models, the key contributor to image quality degradation can be identified, thereby optimizing the design parameters. Notably, when conducting end-to-end system simulations, the error models described in this paper should not be viewed as exhaustive. In future work, we will incorporate additional error models, such as those accounting for optical system attitude corrections, shutter operations, and other factors, into the simulation system. This will enable more accurate calculation and a more precise assessment of the optical image quality of space telescopes.

\normalem
\begin{acknowledgements}

This work was supported by the Special Fund Project for Science and Technology Cooperation Project between Jilin Province and Chinese Academy of Sciences (2024SYHZ0024), and Jilin Provincial Scientific and Technological Development Program(20230201049GX). 

\end{acknowledgements}
  
\bibliographystyle{raa}
\bibliography{bibtex}
 
\end{document}